\documentclass{ws-p8-50x6-00}
\usepackage{makeidx}
\makeindex
\begin{document}
\title{Asymptotic Freedom in Curvature-Saturated Gravity}
\author{S. Capozziello\index{CAPOZZIELLO, S.},
G. Lambiase\index{LAMBIASE, G.}}
 \address{Dipartimento di Scienze Fisiche ``E.R. Caianiello"\\
 Universit\'a di Salerno, 84081 Baronissi (Sa), Italy and\\
Istituto Nazionale di Fisica Nucleare, Sez. di Napoli, Italy\\
E-mail: capozziello@sa.infn.it \qquad  lambiase@sa.infn.it}
\author{Hans-J\"urgen Schmidt\index{SCHMIDT, H.-J.}}
 \address{Freie Universit\"at Berlin, Inst. f. Theor.  Physik,\\
 Arnimallee 14, D-14195 Berlin, Germany and\\
Institut f\"ur Mathematik, Universit\"at Potsdam,\\ Am Neuen
Palais 10, D-14469 Potsdam, Germany\\ E-mail:
hjschmi@rz.uni-potsdam.de}
\maketitle
\abstracts{For a  spatially flat
 Friedmann model\index{Friedmann model}
with line element $ds^2=a^2 [ da^2/B(a)-dx^2-dy^2-dz^2 ] $, the
$00$-component of the Einstein field equation\index{Einstein field
equation}
 reads $8\pi G T_{00}=3/a^2$
containing  no  derivative. For a nonlinear Lagrangian ${\cal
L}(R)$, we obtain  a second--order differential equation for $B$
instead of the expected fourth-order equation. We discuss this
equation for the curvature-saturated model proposed by Kleinert
and Schmidt \cite{kleinert}. Finally, we argue that  asymptotic
freedom $G_{{\rm eff}}^{-1}\to 0$ is fulfilled in
curvature-saturated gravity\index{curvature-saturated gravity}.}
\section{Introduction}
In the past decades, several  extended theories of gravity have
been proposed,  whose effective actions are more general than the
Einstein-Hilbert action\index{Einstein-Hilbert action}. This
approach is motivated by  unification schemes which consider
gravity at  the same level  as  the other interactions of the
elementary particles. In such  theories, we have to define an
effective gravitational coupling\index{effective gravitational
coupling} $G_{\rm eff}$ and an effective cosmological
constant\index{effective cosmological constant} $\Lambda_{\rm
eff}$ which  give the today observed values $G_{\rm
eff}\rightarrow G $ and $\Lambda_{\rm eff}\rightarrow\Lambda$ in
the weak--energy limit.

These extended theories introduce new features into gravitational
 physics which general relativity does not posses, in particular
higher-order terms in curvature invariants or nonminimal couplings
between geometry and scalar fields. For example, {\it asymptotic
freedom}\index{asymptotic freedom} can be related  to
singularity-free cosmological models\index{cosmological models}.
Its emergence could be the result of the fact that gravity is
``induced" by an average effect of the other interactions
\cite{markov,cap}.

However, the concept of gravitational asymptotic freedom is not
completely analogous to that in non--abelian gauge theories of
strong interaction, since a  full quantum  theory of gravity does
not yet exist. So far, asymptotic freedom has been shown for
several classes of gravitational Lagrangians which are not  of
physical interest.

Another interesting feature is that, by extended theories of
gravity,  cosmological singularities can be avoided introducing a
{\it limiting curvature hypothesis} for values of curvature near
the  Planck scale \cite{anini}. This fact is particularly relevant
for string-dilaton gravity\index{string-dilaton gravity} since, by
duality, it allows to recover large classes of cosmological
solutions for $t\rightarrow-\infty$ \cite{vafa}.

 In Ref. \cite{kleinert}, it was argued that the Lagrangian
for gravity should remain bounded at large curvature. One example
for such a behaviour is the curvature-saturated Lagrangian
\begin{equation}\label{1.1}
  {\cal L}_{{\rm CS}}=\frac{1}{16\pi G}\frac{R}{\sqrt{1+l^4R^2}}\,.
\end{equation}
It has been discussed in \cite{kleinert} with methods developed
 for the study of
fourth-order gravity\index{fourth-order gravity} developed in
\cite{schmidt} and \cite{capozziello} and the references cited
therein. In Eq. (\ref{1.1}),  $l$ is a length parameter, and for
$l=0$, ${\cal L}_{{\rm CS}}$ reduces to the Einstein-Hilbert
Lagrangian. Earlier models with a non-linear Lagrangian ${\cal
L}(R)$ made only polynomial approximation like
\begin{equation}\label{1.2}
  {\cal L}=\frac{R}{16\pi G}+\sum_{k=2}^nc_kR^k\,,\quad c_n\neq 0
\end{equation}
sometimes  accompanied by a logarithmic or a $R^m$-term with
non-integer $m$. In all such cases, one had
 $d{\cal L}/{dR}\to \pm
\infty$ for $\vert R \vert \to \infty$. In contrast  to this,
${\cal L}_{{\rm CS}}$ of Eq. (\ref{1.1}) has the behaviour $d{\cal
L}/{dR}\to 0$ for $R\to \pm \infty$.

This limiting property
 can be reformulated in terms  of the effective gravitational
coupling
\begin{equation}\label{1.3}
  G_{{\rm eff}}=\left[16\pi \frac{d{\cal L}}{dR}\right]^{-1}
\end{equation}
as follows: In the previously extended models Eq. (\ref{1.2}),
$G_{{\rm eff}}$ is bounded (including zero),  in contrast to the
curvature-saturated
 model  Eq. (\ref{1.1}) where one gets $\vert
G_{{\rm eff}}\vert \to \infty$ as $\vert R\vert \to \infty$; of
course, for $l=0$ we recover $G_{{\rm eff}}\equiv G$.

Other  limiting behaviours have been discussed  in \cite{cap} and
\cite{markov}. In \cite{cap}, a scalar-tensor theory with
Lagrangian
\begin{equation}\label{1.4}
  {\cal
  L}_{\varphi}=F(\varphi)R+
\frac{1}{2}g^{ij}\varphi_{,i}\varphi_{,j}-V(\varphi)
\end{equation}
has been used where $i, j=0,1,2,3$ and the effective gravitational
constant is
\begin{equation}\label{1.5}
  G_{{\rm eff}}=\frac{1}{16\pi F(\varphi)}
\end{equation}
 instead of  our Eq. (\ref{1.3}). In \cite{markov}, $G_{{\rm eff}}$
was assumed to depend on the matter density in such a way that it
 vanishes for high density, thus leading to asymptotic freedom at
high energies.

In \cite{zhuk}, possible finite--size {\it Casimir}--effects to
the free energy have been calculated for massive and massless
scalar fields, which can produce a quantum-effected effective
gravitational constant.

From a geometric point of view, one may argue as follows: For
large curvatures, the gravitational action $I=\int {\cal L}_{{\rm
C}} \sqrt{-g}d^4x$ shall not depend on any scale; the simplest
Lagrangian leading to such a behaviour is
\begin{equation}\label{1.6}
  {\cal L}_{{\rm C}}=C_{ijkl}C^{ijkl}\,,
\end{equation}
the Lagrangian of the conformally invariant Weyl
gravity\index{Weyl gravity}, see \cite{sch} and the references
cited there.

In the present paper, we extend  the discussion  of Ref.
\cite{kleinert} to  more general types of curvature-saturated
Lagrangians than Eq. (\ref{1.1}), and we give more details about
the set of spatially flat Friedmann models solving the
corresponding field equations. As it will become  clear
 below, the concept of asymptotic freedom is meaningful
  also for curvature-saturated theories and it is
 necessary to define meaningfully the notion of  an effective
cosmological constant.

\section{Deduction of the Field Equations}
For the action $I=\int {\cal L}(R) \sqrt{-g}d^4x$, it is useful
 to write the metric of the expanding spatially flat Friedmann
model as
\begin{equation}\label{2.1}
  ds^2=a^2\left[\frac{da^2}{B(a)}-dx^2-dy^2-dz^2\right]\,,
\end{equation}
with $a>0$ and $B(a)>0$ as shown in \cite{kleinert}. In these
coordinates, the parameter $a$ is called {\it curvature
time}\index{curvature time}
 because the curvature scalar\index{curvature scalar}
has the simple form
\begin{equation}\label{2.2}
  R=-\frac{3}{a^3}\frac{dB}{da}\,,
\end{equation}
which is linear in the only unknown function $B(a)$, and it does
not contain second derivatives. The non-vanishing components of
the Christoffel affinity are as follows
\begin{equation}\label{2.3}
  \Gamma_{00}^0=\frac{1}{a}-\frac{B'}{2B}\,, \quad
  \Gamma_{0\alpha}^{\beta}=\frac{1}{a}\delta_\alpha^\beta\, ,
  \quad
 \Gamma_{\alpha\beta}^0=\frac{B}{a}\delta_{\alpha\beta}\,,
\end{equation}
where the dash denotes $d/da$. The greek  indices assume the
spatial values $\alpha, \beta=1,2,3$, and
$\sqrt{-g}=a^4/\sqrt{B}$. The Ricci tensor\index{Ricci tensor}
 has the components
\begin{equation}\label{2.4}
  R_{\alpha\beta}=\left(\frac{B}{a^2}+\frac{B'}{2a}\right)\delta_{\alpha\beta}\,,
  \quad
  R_{00}=\frac{3}{a^2}-\frac{3B'}{2aB}.
\end{equation}
Together with the metric Eq. (\ref{2.1}),
 its mixed-variant version can be calculated to
\begin{equation}\label{2.5}
  R_{\alpha}^\beta=-\left(\frac{B}{a^4}+\frac{B'}{2a^3}\right)\delta_{\alpha}^{\beta}\,,
  \quad
  R_0^0=\frac{3B}{a^4}-\frac{3B'}{2a^3}\,.
\end{equation}
The most often used direct way to deduce the field equation is to
insert these expressions into the fourth-order field equation
 ${\delta I}/{\delta g_{ij}}=0$ following from the variation of the action $I$.

Here we present a shorter and more direct derivation: We first use
the fact that the field equation also implies ${\delta I}/{\delta
B}=0$, and in a second step,  verify that in spite of this
simplified variation, no spurious solutions appear.

Let us  denote ${d {\cal L}}/{dR}$ by $h(R)$, and ${d^2 {\cal
L}}/{dR^2}$ by $k(R)$. The vacuum field equation reads
 \[
 0=\frac{\delta ({\cal L}\sqrt{-g})}{\delta B}=
 \frac{\partial ({\cal L}\sqrt{-g})}{\partial B}-
 \left(\frac{\partial ({\cal L}\sqrt{-g})}{\partial B'}\right)' \ .
 \]
After multiplying this by $a^7$ to avoid negative $a$-powers, we
obtain
\begin{eqnarray}
 0&=&-a^7{\cal L}(-3B'/a^3)+3a^3(2B-aB')h(-3B'/a^3)+ \nonumber \\
   & &+18B(3B'-aB'')k(-3B'/a^3)\,. \label{2.6}
\end{eqnarray}
It is remarkable that this equation is of second order for one
function $B$ only, but nevertheless, it is equivalent to the whole
fourth-order field equation for the metric (\ref{2.1}). One order
reduction follows  from Eq. (\ref{2.2}), the other  from the fact
that Eq. (\ref{2.6}) is a constraint and not the full dynamical
equation.

To exclude the existence of spurious solutions it now suffices to
see that exactly two free initial conditions can be put: $B(a_0)$
and $B'(a_0)$.

\noindent {\it Example:} Let ${\cal L}=R+2\Lambda$, where
 $\Lambda$ is a constant. Then $h=1$ and $k=0$, and Eq. (\ref{2.6})
reads
\begin{equation}\label{2.7}
  B(a)=\frac{a^4\Lambda}{3}\,.
\end{equation}
The Hubble parameter\index{Hubble parameter}
 $H$ is related to $B$ via
\begin{equation}\label{2.8}
  B(a)=H^2a^4\,,
\end{equation}
(see \cite{kleinert}). So, with Eq. (\ref{2.7}) we get
$\Lambda=3H^2= {\rm constant}$ consistent with the usual de Sitter
space-time\index{de Sitter space-time} calculation.

\section{Solutions of the Field Equations}
Now we consider some more general cases for ${\cal L}(R)$ and look
for the corresponding solutions of Eq. (\ref{2.6}).

\subsection{Solution for Einstein's theory}
To get a feeling for the coordinates (\ref{2.1}) we first look for
the $00-$component  of the Einstein field equation
 \[
 8\pi GT_{ij}=R_{ij}-\frac{1}{2}Rg_{ij}\,.
 \]
Using Eqs. (\ref{2.2}) and (\ref{2.4}), we obtain
\begin{equation}\label{3.1}
  8\pi GT_{00}=\frac{3}{a^2}\,.
\end{equation}
 The  energy density is always non-negative for
spatially flat Friedmann models in Einstein's theory. Eq.
(\ref{3.1}) provides us with another justification for calling
 this  coordinate
curvature-coordinate: $a$ is chosen such that $T_{00}$ does not
 depend at all on the function $B(a)$.

Raising one index
 of Eq. (\ref{3.1}) we find
\begin{equation}\label{3.2}
  8\pi G\rho\equiv 8\pi GT_0^0=\frac{3B}{a^4}\,.
\end{equation}
Together with Eq. (\ref{2.8}), this  represents a good consistency
test: it yields the Friedmann equation
\begin{equation}\label{3.3}
  8\pi G\rho=3H^2\, ,
\end{equation}
i.e., curvature time and synchronous time give identical results.

For the equation of state $p=\alpha\rho$ we have
\begin{equation}\label{3.4}
  \rho a^{3(1+\alpha)}= {\rm constant}\,.
\end{equation}
Together with Eq. (\ref{3.2}), this yields
\begin{equation}\label{3.5}
  B=B_0a^{1-3\alpha}\,,
\end{equation}
with a positive constant $B_0$. This is consistent with Eq.
(\ref{2.7}) for $\alpha=-1$.

\subsection{Solution for ${\cal L}=R^m$.}
For the Lagrangian ${\cal L}=R^m$, with constant $m$ ($m\neq
0,1$), Eq. (\ref{2.6}) simplifies to
\begin{eqnarray}
  0&=&2m(3m-4)BB'+(m-1)aB^{'\,2} \nonumber \\
  & & -2m(m-1)aBB''\,. \label{3.6}
\end{eqnarray}
We insert
\begin{equation}\label{3.7}
  B=\exp{\int \beta da}\,.
\end{equation}
The integration constant in Eq. (\ref{3.7}) need not be specified,
because Eq. (\ref{3.6}) implies that with $B(a)$ also $CB(a)$,
with any positive constant $C$, represents a solution.

Eq. (\ref{3.7}) implies $B'=\beta B$ and $B''=(\beta'+\beta^2)B$.
Then Eq. (\ref{3.6}) becomes
\begin{eqnarray}
  0&=&2m(3m-4)\beta-(m-1)(2m-1)a\beta^2 \nonumber \\
    & & -2m(m-1)a\beta' \,. \label{3.8}
\end{eqnarray}
Here we insert $\beta=\gamma/a$, $\beta'=\gamma'/a-\gamma/a^2$ and
get
\begin{eqnarray}
  0&=&2m(4m-5)\gamma-(m-1)(2m-1)a\gamma^2 \nonumber \\
    & & -2m(m-1)a\gamma' \,. \label{3.9}
\end{eqnarray}
By putting $x=\ln a$, Eq. (\ref{3.9}) can be rewritten in the form
\begin{equation}\label{3.10}
  \frac{d\gamma}{dx}=\gamma\left[\frac{4m-5}{m-1}-\frac{2m-1}{2m}\gamma\right]\,.
\end{equation}
The de Sitter space-time is represented by $\gamma\equiv 4$. From
Eq. (\ref{3.10}), it becomes clear that it is a solution for $m=2$
only, i.e. for ${\cal L}=R^2$.

All solutions of (\ref{3.10}) can be given in closed form: For
$m=5/4$ we get with any constant $x_0$: $$
\gamma=\frac{5}{3(x-x_0)}\ , $$ and together with $a_0=e^{x_0}$
and Eq. (\ref{3.7}) finally
\begin{eqnarray}
 B(a)&=&\exp{\left[\frac{5}{3}\int\frac{da}{a\ln (a/a_0)}
 \right]}= \nonumber \\
  &=&B_0\left(\ln\frac{a}{a_0}\right)^{5/3}\,.
\label{3.11}
\end{eqnarray}
For $m=1/2$ we find  $\displaystyle{\gamma=\pm e^{6(x-x_0)}}$ and
\begin{equation}\label{3.12}
  B(a)=\exp[\pm (a/a_0)^6]\,.
\end{equation}
Let us study the stability of these solutions; instead of directly
comparing with  neighboring functions $B(a)$
 it is easier to consider the
neighborhood of the constant solution: $\gamma= {\rm  constant}$
in Eq. (\ref{3.10}).  Up to the uninteresting solution $\gamma=0$
representing $R\equiv 0$ we find
 \[
 \gamma=\gamma_0\equiv \frac{4m-5}{m-1}\frac{2m}{2m-1},
 \]
 i.e.
\begin{equation}\label{3.13}
  \frac{d\gamma}{dx}=\gamma (\gamma_0-\gamma)\frac{2m-1}{2m}\,.
\end{equation}
For $m>1/2$ this implies  stability. For $m=2$ this proves again
the attractor\index{attractor} property of the de Sitter
space-time.

In synchronized time, power-law inflation\index{power-law
inflation} is described by $a(t)\sim t^n$, $n\geq 1$. In our
coordinates, such solutions correspond to a constant $\gamma$
within the interval $2\leq \gamma < 4$.

\subsection{Solutions for Curvature-Saturated Lagrangians}
As  another model, consider the
 high-curvature ansatz
\begin{equation}\label{3.14}
  {\cal L}=\Lambda+\frac{C}{R}\,,
\end{equation}
 in which
 $h(R)=-C/R^2$, $k(R)=2C/R^3$, with constants $\Lambda$ and
$C$. The ansatz (\ref{3.14})
 is supposed to approximate
 the regions $R\to \pm
\infty$, and it will be matched at small curvature values to a
polynomial Lagrangian of type Eq. (\ref{1.2}). Here we are only
interested in the high-curvature regions. The concrete values of
$\Lambda$ and $C$ are not yet fixed, and they may be different for
$R\to +\infty$ and $R\to -\infty$, respectively.

For $\Lambda=0$ we can directly apply Eq. (\ref{3.13}) with
$m=-1$, i.e., $\gamma_0=3$,  and
\begin{equation}\label{3.15}
 \frac{d\gamma}{dx}=\frac{3}{2}\gamma(3-\gamma)\,.
\end{equation}
This equation can be integrated in closed form, but we only need
that $\gamma=3$ represents power-law inflation and that this
represents an attractor solution. In synchronized coordinates, it
reads
\begin{equation}\label{3.16}
  ds^2=dt^2-t^4(dx^2+dy^2+dz^2)\,.
\end{equation}
It gives $R\to 0$ as $t\to \infty$. Thus, for sufficiently large
$t$, the development leads   to our  Universe today.

For $\Lambda\neq 0$, we insert (\ref{3.14}) into (\ref{2.6}), and
have to solve
\begin{equation}\label{3.17}
 3\Lambda (B')^3=2a^2C[5BB'+aB^{'\, 2}-2aBB'']\,.
\end{equation}
This equation will be integrated further in future  work.

\section{Curvature-Saturated Gravity with Matter}
To include matter, we have to replace the l.h.s. of Eq.
(\ref{2.6}) by $a^7\rho$. For ideal fluid, we can use Eq.
(\ref{3.4}) as an  equation of state. As a test of this procedure,
we can insert ${\cal L}=R/16\pi G$, and get
$\displaystyle{a^7\rho=\frac{3a^3B}{8\pi G}}$, which is consistent
with Eq. (\ref{3.2}).

Therefore, matter with equation of state $p=\alpha\rho$ can be
described by $\rho=\rho_0a^{-3(1+\alpha)}$ and the equations of
motion are
\begin{eqnarray}
 a^7\rho&=&-a^7{\cal L}(R)+3a^3(2B-aB')h(R)  \nonumber \\
 & & + 18B(3B'-aB'')k(R)\,,
\label{4.1} \\
 R&=&-3B'a^{-3}\,. \nonumber
\end{eqnarray}
We stress that such equations hold for general
 ${\cal L}(R)$--Lagrangians and for any
perfect fluid matter\index{perfect fluid matter}.

\section{Conclusions}

In this paper, we continued the discussion on properties of the
curvature-saturated cosmological models proposed in Ref.
\cite{kleinert}.
 Especially, we deduced the main equation {\it with} matter, Eq.
(\ref{4.1}), for a spatially flat Friedmann model in the curvature
coordinates (\ref{2.1}).

We have shown that from the singularity\index{singularity}, the
Universe\index{Universe} expands via power-law inflation (which
represents a
 transient attractor\index{transient attractor})
 to
the actual state.

In order to find the Wheeler-de Witt equation\index{Wheeler-de
Witt equation} for fourth--order gravity models (see Refs.
\cite{schmidt,fabris,sch1} and the papers cited there), one
usually has to introduce, in a more or less natural way, more
degrees of freedom to reduce the equation down to second order.
Here we presented a version where the Lagrangian can be directly
used: ${\cal L}(R) \sqrt{-g}$ depends on $a$, $B$ and $B'$ only;
of course, differently from the use of the synchronized time
coordinate, we have now an explicit dependence of the metric
coefficients  on  the time-like coordinate $a$.

As a final remark, we see that asymptotic freedom $G_{\rm
eff}^{-1}=16\pi d{\cal L}/dR\rightarrow 0$ can  easily be
 incorporated  also in curvature-saturated gravity but its meaning is
different from that in \cite{cap}. There $G_{\rm eff}$ is a
function of matter density which regulates the gravitational
coupling, here
 it is the scalar curvature (i.e. the form of the gravitational
Lagrangian) which leads  towards  the saturation and then towards
asymptotic freedom. Future studies will be devoted to more
physically motivated effective Lagrangians.

\section*{Acknowledgment}
We thank Professor Hagen Kleinert for his constant encouragement
over the
 years. He combines horizontal broadness  in form of a large number of
topics where he is really interested in with vertical deepness
shown by
 his ability to criticize weaknesses in  manuscripts to which
 even specialists in the field did not become aware.

H.-J.S. thanks the colleagues of Salerno University where this
work has been completed, especially Professor Gaetano Scarpetta,
for kind hospitality. The researches of  S.C. and G.L.  have been
supported by fund MURST PRIN 99. H.-J.S. is supported by the HSP
III-program Potsdam.


\begin{thebibliography}{99}
%
\bibitem{kleinert} H. Kleinert\index{KLEINERT, H.}
and H.-J. Schmidt\index{SCHMIDT, H.-J.}, {\em Cosmology
                   with curvature-saturated Lagrangians}, e-print: gr-qc/0006074.
\bibitem{markov} M. Markov\index{MARKOV, M.}, {\em
 Physics-Uspekhi} {\bf 37}, 57 (1994).
\bibitem{cap}S. Capozziello\index{CAPOZZIELLO, S.} and
R. de Ritis\index{DE RITIS, R.}, {\em  Phys.
             Lett. A} {\bf  208}, 181 (1995);\\
             S. Capozziello\index{CAPOZZIELLO, S.},
 R. de Ritis\index{DE RITIS, R.}, and A. Marino\index{MARINO, A.},
 {\em Phys.   Lett. A} {\bf 249}, 395 (1998).
\bibitem{anini} Y. Anini\index{ANINI, Y.},
p.183 in {\em Current topics in mathematical cosmology},
                Eds.: M. Rainer\index{RAINER, M.},
H.-J. Schmidt\index{SCHMIDT, H.-J.} (WSPC Singapore
                1998);\\ R. Brandenberger\index{BRANDENBERGER, R.},
V. Mukhanov\index{MUKHANOV, V.} and A.
                Sornborger\index{SORNBORGER, A.},
 {\em Phys. Rev. D} {\bf  48}, 1629 (1993).
\bibitem{vafa} A.A. Tseytlin\index{TSEYTLIN, A. A.} and
C. Vafa\index{VAFA, C.}, {\em Nucl. Phys. B} {\bf 372}, 443
(1992).
\bibitem{schmidt} H.-J. Schmidt\index{SCHMIDT, H.-J.},
 {\em Phys. Rev. D} {\bf  49}, 6354    (1994); e-print: gr-qc/9404038.
\bibitem{capozziello} S. Capozziello\index{CAPOZZIELLO, S.},
G. Lambiase\index{LAMBIASE, G.}, and H.-J.
                   Schmidt\index{SCHMIDT, H.-J.},
 {\em Ann. Phys. (Leipz.)} {\bf 9}, 39 (2000).
\bibitem{zhuk} A. Zhuk\index{ZHUK, A.} and
H. Kleinert\index{KLEINERT, H.}, {\em Theor. and Math. Phys.}
               {\bf 109}, 1483 (1996);
 H. Kleinert and A. Zhuk: unpublished Preprint 1993: {\em
 Finite-Size and Temperature properties of
 Matter and Radiation fluctuations in Closed Friedmann Universe};
 H. Kleinert and  A. Zhuk: {\em Casimir
effect at nonzero temperature in closed Friedmann universe}; the
files can be downloaded
 from  http://www.physik.fu-berlin.de/\~{}kleinert/ as  Nr. 218 and
251 resp.
\bibitem{sch} V. Dzhunushaliev\index{DZHUNUSHALIEV, V.}
 and H.-J. Schmidt\index{SCHMIDT, H.-J.}, {\em
 J. Math. Phys.} {\bf 41}, 3007 (2000); e-print:
              gr-qc/9908049.
\bibitem{fabris} J. Fabris\index{FABRIS, J.} and S. Reuter\index{REUTER, S.},
 {\em Gen. Rel. Grav.} {\bf  32}, 1345 (2000).
\bibitem{sch1} M. Bachmann\index{BACHMANN, M.}
 and H.-J. Schmidt\index{SCHMIDT, H.-J.},
{\em Phys. Rev. D} {\bf  62}, 043515 (2000); e-print:
gr-qc/9912068.
\end{thebibliography}
\end{document}